\documentclass[]{aa} % for the letters 
\usepackage[]{natbib}
\usepackage{graphicx,color}
\usepackage{txfonts}
\usepackage[colorlinks=true, citecolor=blue]{hyperref}
\usepackage[table]{xcolor}
\usepackage{multirow} %multiple row in tables
\usepackage{boldline} %bold lines in tables
\usepackage{adjustbox}
\usepackage{float}
\usepackage{booktabs}

\newcommand{\orcid}[1]{\href{https://orcid.org/#1}{\includegraphics[width=10pt]{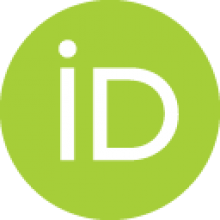}}}

\begin{document} 

\title{APOGEE-2S Mg-Al anti-correlation of the metal-poor globular cluster NGC~2298}

\author{
        Ian Baeza\inst{1}\thanks{ibaeza2017@udec.cl}\orcid{0000-0002-9881-6336}
        \and
        Jos\'e G. Fern\'andez-Trincado\inst{2}\thanks{jose.fernandez@ucn.cl}\orcid{0000-0003-3526-5052}
        \and
        Sandro Villanova\inst{1}\thanks{svillanova@astro-udec.cl}\orcid{0000-0001-6205-1493}
        \and 
        Doug Geisler\inst{1,3,4}
        \and
        Dante Minniti\inst{5,6}\orcid{0000-0002-7064-099X}
        \and 
        Elisa R. Garro\inst{5}\orcid{0000-0002-4014-1591}
        \and 
        Beatriz Barbuy\inst{7}\orcid{0000-0001-9264-4417}
        \and 
        Timothy C. Beers\inst{8}\orcid{0000-0003-4573-6233}
        \and
        Richard R. Lane\inst{9}\orcid{0000-0003-1805-0316}
%        \and 
%      Baitian Tang\inst{10}\orcid{0000-0002-0066-0346}
      }

        \authorrunning{Ian Baeza et al.} 
        
\institute{
        Departamento de Astronom\'ia, Casilla 160-C, Universidad de Concepci\'on, Concepci\'on, Chile
            \and 
         Instituto de Astronom\'ia, Universidad Cat\'olica del Norte, Av. Angamos 0610, Antofagasta, Chile
         \and
         Departamento de Astronom\'ia, Universidad de La Serena, 1700000 La Serena, Chile
         \and
         Instituto de Investigaci\'on Multidisciplinario en Ciencia y Tecnolog\'ia, Universidad de La Serena. Benavente 980, La Serena, Chile
         \and
         Depto. de Cs. F\'isicas, Facultad de Ciencias Exactas, Universidad Andr\'es Bello, Av. Fern\'andez Concha 700, Las Condes, Santiago, Chile
         \and
         Vatican Observatory, V00120 Vatican City State, Italy
         \and 
         Universidade de S\~ao Paulo, IAG, Rua do Mat\~ao 1226, Cidade Universit\'aria, S\~ao Paulo 05508-900, Brazil
         \and 
         Department of Physics and JINA Center for the Evolution of the Elements, University of Notre Dame, Notre Dame, IN 46556, USA
         \and
         Centro de Investigaci\'on en Astronom\'ia, Universidad Bernardo O'Higgins, Avenida Viel 1497, Santiago, Chile
%         \and
%         School of Physics and Astronomy, Sun Yat-sen University, Zhuhai 519082, China  
    }

        \date{Received ...; Accepted ...}
        \titlerunning{NIR chemical abundances in NGC~2298}
        
        % \abstract{}{}{}{}{} 
        % 5 {} token are mandatory
        
        \abstract
        {
We present detailed elemental abundances and radial velocities of stars in the metal-poor globular cluster (GC) NGC~2298, based on near-infrared high-resolution ($R\sim$ 22,500) spectra of 12 members obtained during the second phase of the Apache Point Observatory Galactic Evolution Experiment (APOGEE-2) at Las Campanas Observatory as part of the seventeenth Data Release (DR~17) of the Sloan Digital Sky Survey IV (SDSS-IV). We employed the Brussels Automatic Code for Characterizing High accuracy Spectra (\texttt{BACCHUS}) software to investigate abundances for a variety of species including $\alpha$ elements (Mg, Si, and Ca), the odd-Z element Al, and iron-peak elements (Fe and Ni) located in the innermost regions of NGC~2298. We find a mean and median metallicity [Fe/H] $ = -1.76$ and $-1.75$, respectively, with a star-to-star spread of 0.14 dex, which is compatible with the internal measurement errors.  Thus, we find no evidence for an intrinsic [Fe/H] abundance spread in NGC~2298. The typical $\alpha-$element enrichment in NGC~2298 is overabundant relative to the Sun, and it follows the trend of other metal-poor GCs. We confirm the existence of an Al-enhanced population in this cluster, which is clearly anti-correlated with Mg, indicating the prevalence of the multiple-population phenomenon in NGC~2298. 
}
        
        \maketitle
        
\section{Introduction}
\label{Introduction}

Globular clusters (GCs) are important long-lived time capsules that provide information about the primordial evolutionary stages of their host galaxies. For several decades, GC stars have been known to display a zoo of peculiarities, such as light-element \citep[][and references therein]{Carretta2009, Pancino2017, Schiavon2017, Masseron2019, Meszaros2020, Meszaros2021, Geisler2021} and heavy-element \citep{Carretta2021, Marino2021, Trincado2022, Trincado2021} abundance variations over a wide range of metallicities, which have been attributed to the 
multiple-populations (MPs) phenomenon \citep[for a thorough review, see][]{Bastian2018}.

The southern cluster NGC~2298, located in the constellation Puppis, has long been recognized to be among the most metal-poor GCs in the Milky Way (MW) \citep{Hesser1985, McWilliam1992}. However, the metallicity estimates reported for this object have covered a range from [Fe/H]$= -1.96$ to $-1.71$ \citep[see e.g.,][]{Frogel1983, Cohen1983, Zinn1984, Zinn1985, McWilliam1992, Geisler1995, Salaris1996, Carretta1997, Kraft2003, Pritzl2005, Carretta2009a, Carrera2013, Roediger2014, Yong2014}, creating some uncertainty regarding its status as one of the most metal-poor systems. This cluster is also well known for its lack of a clear indication of MPs along the main sequence (MS) \citep{Piotto2015}, as well as likely along the horizontal branch (HB) population \citep{Rani2021}, in near-UV/optical color-magnitude diagrams (CMDs). However, its red-giant branch (RGB) displays clear evidence for both first- and second-generation stars  \citep[e.g.,][]{Milone2017}.

NGC~2298 is a particularly interesting GC, as its origin and nature still remains controversial. For instance, some studies have proposed that it is likely associated with the Monoceros progenitor galaxy \citep{Crane2003, Martin2004, Forbes2010}; however, more recent studies conducted by \citet{Massari2019} and \citet{Malhan2022} have linked this GC to the \textit{Gaia}-Sausage/Enceladus merger \citep{Belokurov2018, Helmi2018Natur}. A few previous and more recent studies have also suggested the presence of extra-tidal features around NGC~2298 \citep{Leon2000, Balbinot2011, Carballo-Bello2018, Piatti2020, Sollima2020, Ibata2021}; however, their existence has been questioned based on more recent investigations of deep imaging from the Dark Energy Camera \citep{Zhang2021}.  

While NGC~2298 has been widely studied photometrically (in the optical and UV), detailed spectroscopic information about it remains sparse. Furthermore, spectroscopic evidence for the existence of multiple stellar populations in this GC has not been reported in the literature yet. Located at 15.1 kpc \citep{Baumgardt2021} from the Galactic center, it is an old \citep[$\sim12 -- 13$ Gyr;][]{Monty2018} GC that lies in a region of relatively low foreground interstellar reddening, with E(B$-$V) $\sim 0.14$ -- $0.16$ \citep{Kraft2003, Piotto2015, Monty2018}, making it an excellent candidate for the study of MPs in GCs at a low Galactic latitude ($b\sim-16^{\circ}$). In the present work, we report the first high-resolution near-infrared (NIR) spectral study of elemental-abundance estimates for a variety of chemical species for a sample of 12 sources in the innermost regions of NGC~2298 and we characterize its chemical composition. For the first time, we report information on the main element families, namely the $\alpha$ elements (Mg, Si, Ca), the odd-Z element Al, and the Fe-peak elements (Fe and Ni). 

The structure of this paper is as follows. Section \ref{Data} describes the data, and Section \ref{Sample} describes the identified cluster members. Sections \ref{atmospheric} and \ref{abundances} describe the derivation of atmospheric parameters and elemental abundances for the individual stars. Section \ref{Results} presents our abundance analysis, and Section \ref{conclusions} provides our concluding remarks.

\section{Data}
\label{Data}

The Apache Point Observatory Galactic Evolution Experiment II survey \citep[APOGEE-2;][]{Majewski2017} is one of the internal programs of the Sloan Digital Sky Survey-IV \citep[SDSS-IV;][]{Blanton2017}.\ It was developed to provide precise radial velocities (RV errors $<$ 1 km s$^{-1}$) and detailed chemical abundances for an unprecedented large sample of giant stars, aiming to unveil the dynamical structure and chemical history across the entire MW. The APOGEE-2 instruments (capable of observing up to 300 objects simultaneously) are high-resolution ($R\sim22,500$), NIR spectrographs \citep{Wilson2019} observing all the components of the MW (halo, disk, and bulge) from the Northern Hemisphere on the 2.5m telescope at the Apache Point Observatory \citep[APO, APOGEE-2N;][]{Gunn2006} and the Southern Hemisphere on the Ir\'en\'ee du Pont 2.5m telescope at the Las Campanas Observatory \citep[LCO, APOGEE-2S;][]{Bowen1973}. Each instrument records most of the \textit{H} band (1.51$\mu$m -- 1.69$\mu$m) on three detectors, with coverage gaps between $\sim$1.58--1.59$\mu$m and $\sim$1.64--1.65$\mu$m, and with each fiber subtending a $\sim$2'' diameter on-sky field of view in the northern instrument and 1.3'' one in the southern instrument.

DR~17 \citep{Abdurro2021} is the final release of the APOGEE-2 survey from SDSS-IV. It includes all APOGEE-2 data taken at APO through November 2020 and at LCO through January 2021. The dual APOGEE-2 instruments have observed more than 700,000 stars throughout the MW. We refer the reader to \citet{Zasowski2017}, \citet{Beaton2021}, and \citet{Santana2021} for further details regarding the targeting strategy of the APOGEE-2 survey.
  
Spectra for this project were reduced as described in \citet{Nidever2015}; they were also analyzed using the APOGEE Stellar Parameters and Chemical Abundance Pipeline \citep[ASPCAP;][]{Garcia2016}, and the libraries of synthetic spectra described in \citet{Zamora2015}. The accuracy and precision of the atmospheric parameters and chemical abundances are extensively analyzed in \citet{Holtzman2018}, \citet{Henrik2018}, and \citet{Henrik2020}, while details regarding the custom \textit{H}-band line list are fully described in \citet{Shetrone2015}, \citet{Hasselquist2016}, \citet{Cunha2017}, and \citet{Smith2021}. 

\section{Sample}
\label{Sample}

The metal-poor GC NGC~2298 was included in the targeting strategy of the APOGEE-2S program at the end of the survey. Here we analyze for the first time the APOGEE-2S spectra for potential members of NGC~2298 -- centered on $\alpha =$ 06:48:59.41 and $\delta =$ $-$36:00:19.1 -- observed from LCO by the APOGEE-2S survey in January 2021. 

\begin{figure}
        \centering
        \includegraphics[width=0.5\textwidth]{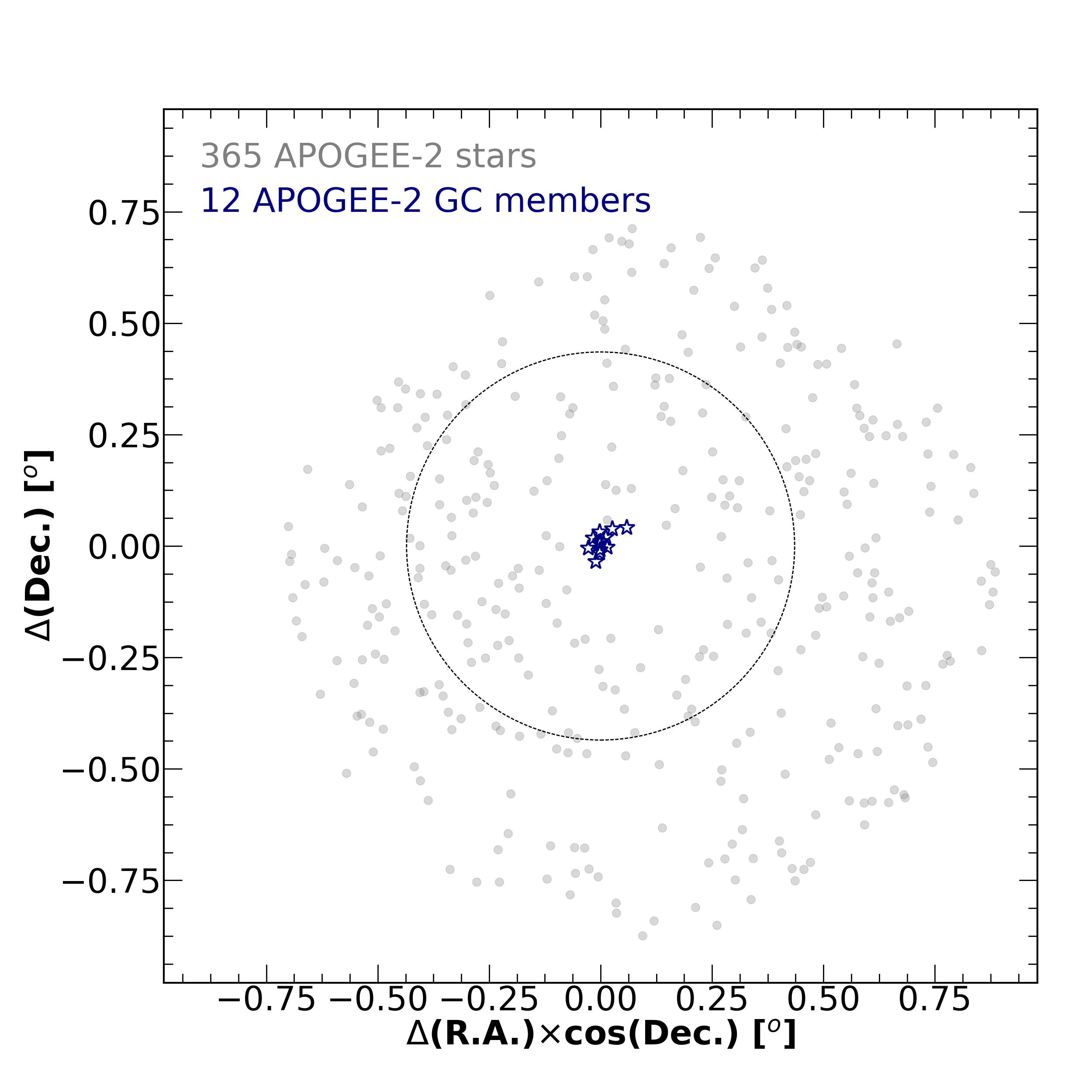}
        \caption{Spatial distribution of sources (gray dots) surveyed by APOGEE-2S toward the NGC~2298 field. Potential cluster members analyzed in this work are marked with empty, navy blue "star" symbols. The large black dashed circle highlights the cluster tidal radius, $r_{\rm t}= 26.12$ arcmin at d$_{\odot} = $9.83$\pm$0.17 kpc \citep{Baumgardt2021}. The image is oriented with the east pointing to the right and north pointing up.}
        \label{Figure1}
\end{figure}

The APOGEE-2S plug-plates (PlateID $=$ 12836 and 12837), containing the NGC~2298 field (LocID $=$ 7253), are centered on ($\alpha$, $\delta$) $\sim$ (102.35549$^{\circ}$, $-$36.0852$^{\circ}$) and contain 365 science fibers. From these, 12 stars lie very close to the cluster center, which is well within the tidal radius (r$_{\rm t} \lesssim $ 26.12 arcmin)\footnote{\url{https://people.smp.uq.edu.au/HolgerBaumgardt/globular/parameter.html}} at d$_{\odot} = 9.83 \pm 0.17$ kpc \citep{Baumgardt2021}, as shown in Figure \ref{Figure1}. 

\begin{figure*}
        \centering
        \includegraphics[width=1.0\textwidth]{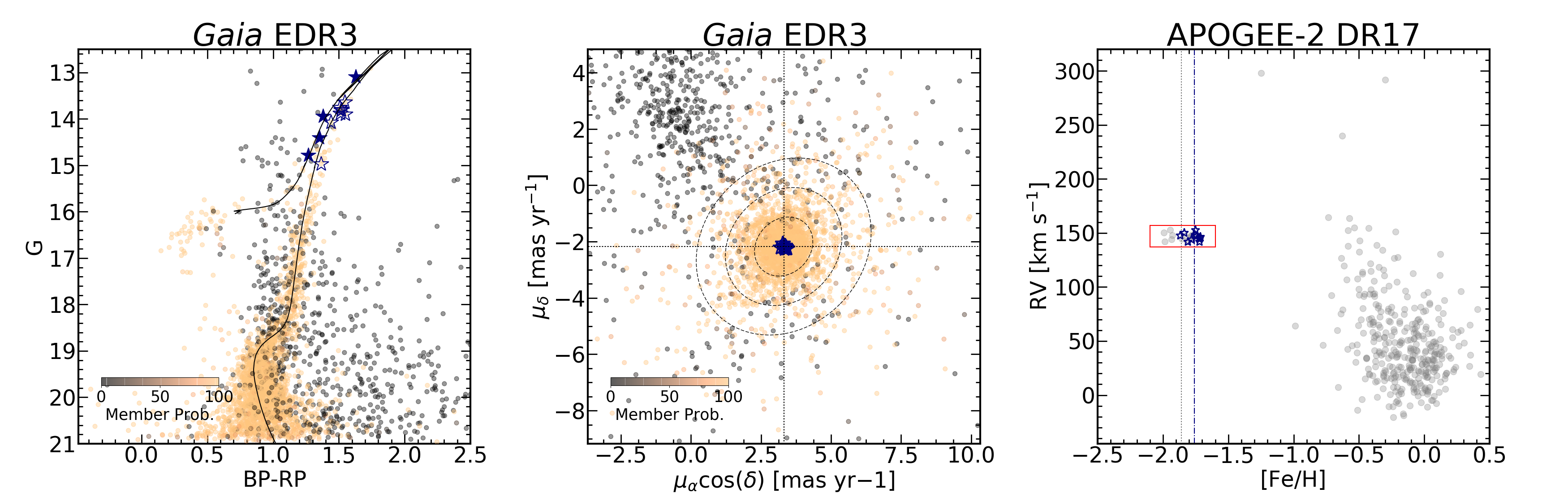}
        \caption{Main physical properties of NGC~2298 stars. Left panel:  \textit{Gaia} optical CMD with membership probabilities color-coded as shown in the inset. RGB and AGB stars are denoted by open and filled symbols, respectively. A \texttt{PARSEC} isochrone of 12 Gyr and a metallicity of $-1.75$ is also overplotted with black solid lines. Middle panel: \textit{Gaia} EDR3 vector point diagram (VPD). The three concentric dashed ellipses represent the 1-, 2-, and 3-$\sigma$ level of the highest likelihood cluster members with $>90\%$ membership probability. The two black dashed lines are shown for visual aids and show the nominal proper motions of the cluster taken from \citet{Vasiliev2021}. Right panel: Metallicity ([Fe/H]) versus radial velocity for 365 APOGEE-2S stars toward the NGC~2298 field. \texttt{ASPCAP} results are denoted by gray dots, while the \texttt{BACCHUS} results are denoted by navy blue star symbols for the 12 cluster members. The red box, which is arbitrarily limited between [Fe/H] $=$ $-2.1$ to $-1.6$, centered on RV$= 147.15 $ km s$^{-1}$ according to \citet{Baumgardt2021}, and limited between $\pm$10 km s$^{-1}$, encloses the highest likelihood cluster members with \texttt{ASPCAP} (gray dots) and \texttt{BACCHUS} (navy blue stars) determinations. The gray dotted and navy blue dash-dotted lines mark the average $\langle$[Fe/H]$\rangle$ of the 12 NGC~2298 stars analyzed in this work, and they are centered at $\langle$[Fe/H]$\rangle$ $= -1.86$ (\texttt{ASPCAP} results) and  $\langle$[Fe/H]$\rangle$ $= -1.76$ (\texttt{BACCHUS} results), respectively. The 12 stars examined in this work are denoted by navy blue star symbols in all panels. }
        \label{Figure2}
\end{figure*}

Figure \ref{Figure2} shows  that the 12 sources analyzed in this work share common properties making them very likely NGC~2298 members, that is to say their position along the main branches of the cluster CMD in the \textit{Gaia} EDR3 bands; nominal \textit{Gaia} EDR3 proper motions, ($\mu_{\alpha}\cos{}\delta$; $\mu_{\delta} $) $=$ (3.320$\pm$0.025; $-$2.175 $\pm$ 0.026) mas yr$^{-1}$; radial velocity, RV$= 147.15 \pm 0.57$ km s$^{-1}$ \citep{Vasiliev2021}; and metallicity. The same figure (\textit{left and middle panel}) also shows the membership probability provided by \citet{Vasiliev2021} for the NGC~2298 field. The 12 target stars occupy the region of the CMD dominated by highest likelihood cluster stars on the red giant branch (RGB) and asymptotic giant branch (AGB).

The APOGEE-2S spectra used in this work have a typical signal-to-noise ratio (S/N) larger than 109 pixel$^{-1}$, which was reached from two (ten stars) and four (two stars) field visits, making these spectra ideal to obtain reliable elemental abundances for several chemical species accessible from the \textit{H} band in the metal-poor regime. Additionally, we find the typical radial velocity scatter (\texttt{VSCATTER}) to be less than 0.3 km s$^{-1}$ for all sources in our sample, with no strong evidence for radial velocity variations that might affect the resulting elemental abundances. 

Figure \ref{Figure2} also shows the radial velocity against the metallicity for the 365 sources surveyed by APOGEE-2S in the region of this GC. \texttt{ASPCAP}--[Fe/H] determinations are highlighted by gray open symbols, while the \texttt{BACCHUS}--[Fe/H] ones (see Section \ref{abundances}) are marked with navy blue symbols. This figure reveals that \texttt{BACCHUS}--[Fe/H] is systematically offset by roughly 0.1 dex to higher metallicity compared to the \texttt{ASPCAP}--[Fe/H] abundance ratio, which is most likely due to systematics between \texttt{BACCHUS} and \texttt{ASPCAP},  similar to other studies of the APOGEE-2 spectra \citep[see, e.g.,][]{Masseron2019, Fernnadez-Trincado2020_Aluminum,Meszaros2020}.

\section{Atmospheric parameters}
\label{atmospheric}

Adopting the typical value $R_{V}=$ 3.1, E(B$-$V) $=$ 0.14 \citep{Monty2018} and a distance of 9.83 kpc \citep{Baumgardt2021}, we first de-reddened the optical \textit{Gaia} EDR3 photometric bands. We obtain T$_{\rm eff}$ and $\log$ (\textit{g}) from photometry by determining the reddening-corrected CMD of NGC~2298.  We then horizontally projected the position of each observed star until it intersected the \texttt{PARSEC} \citep{Bressan2012} isochrone, chosen to have an age of $\sim$12 Gyr \citep{Monty2018}, a metallicity of [Fe/H] $\sim -1.75$ (see Table \ref{table1}), and an assumed T$_{\rm eff}$ and $\log$ (\textit{g}) to be the temperature and gravity at the point of the isochrones that has the same optical magnitude as the star. Since no differential reddening was present, we could easily separate RGB from AGB stars. For each group, we used the corresponding part of the isochrone to obtain the parameters, that is, we used the isochrone RGB to obtain T$_{\rm eff}$ and $\log$ (\textit{g}) for RGB stars, and the isochrone AGB to obtain T$_{\rm eff}$ and $\log$ (\textit{g}) for AGB stars. Finally, microturbulence velocities $\xi_{t}$ were determined empirically with the \texttt{BACCHUS} code \citep{Masseron2016}. The resulting atmospheric parameters and elemental abundances for the 12 stars examined in the innermost regions of NGC~2298 are listed in Table \ref{table1}. 

\begin{figure}
        \centering
        \includegraphics[width=0.5\textwidth]{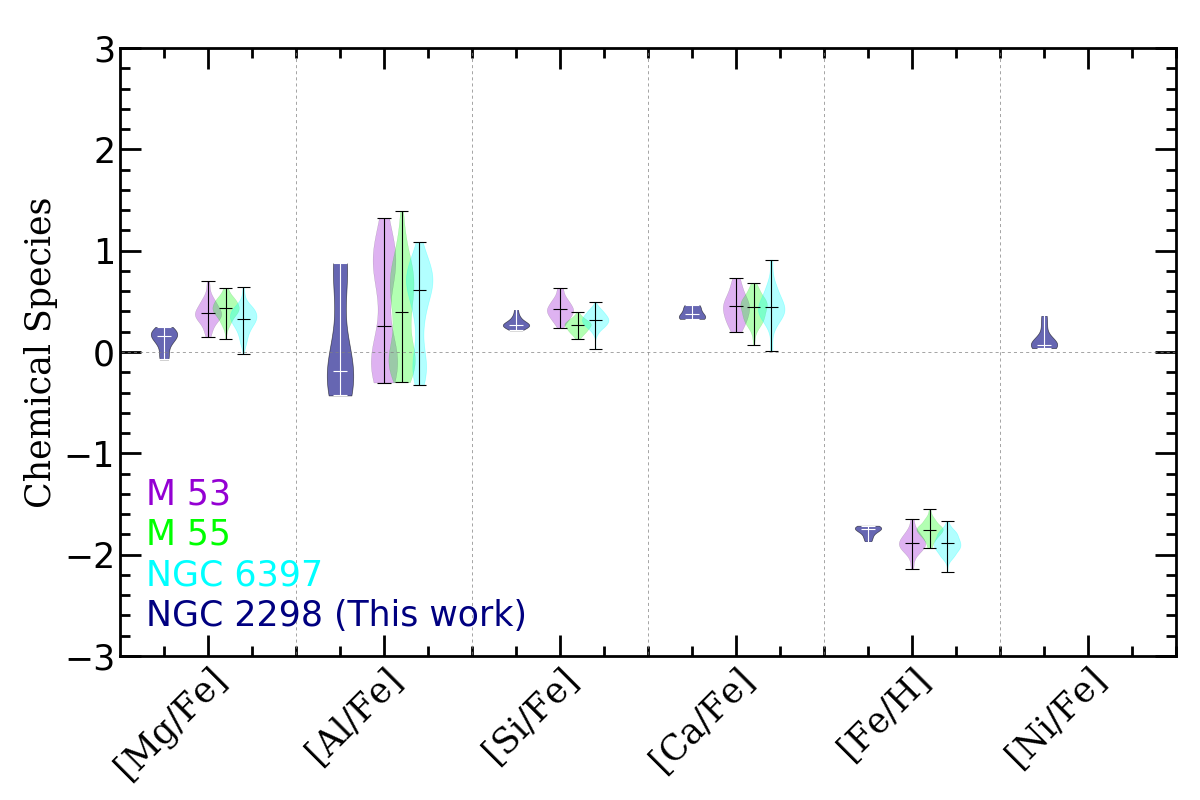}
        \caption{[X/Fe] and [Fe/H] abundance kernel density estimation comparison between NGC~2298 (navy blue symbols on the left) and cluster stars (violet, lime green, and cyan symbols for M~53, M~55, and NGC~6397, respectively, on the right) taken from \citet{Meszaros2020}. Each violin-type shape indicates the median and limits of the distribution with horizontal lines.}
        \label{Figure3}
\end{figure}

\section{Abundance determinations}
\label{abundances}

Once the final atmospheric parameters were computed, abundances for a variety of elements could be derived line-by-line using the \texttt{BACCHUS} code. Thus, chemical abundances were derived for six atomic species (Mg I, Al I, Si I, Ca I, Fe I, and Ni I) from a local thermodynamics equilibrium (LTE) analysis using \texttt{BACCHUS} combined with the \texttt{MARCS} model atmospheres \citep{Gustafsson2008}, and following the same technique as described in several APOGEE-2 works \citep[see, e.g.,][for instance]{FT_NGC6723, FT_NGC6522}, and summarized here for guidance. With the atmospheric parameters determined in Section \ref{atmospheric}, the first step consisted of determining the metallicity from selected Fe I lines, the micro-turbulence velocity ($\xi_{t}$), and the convolution parameter. Other chemical species not reported in this work were found to have very weak lines and/or were too heavily blended by telluric features to provide reliable abundances. 

With the metallicity and main atmospheric parameters fixed, we then computed the abundance of each chemical species as follows: (a) We performed a synthesis using the full set of atomic and molecule line lists fully described in \citet{Smith2021}. This set of lines is internally labeled as \texttt{turbospec.20180901t20.atoms} and \texttt{turbospec.20180901t20.molec} based on the date of creation in the format YYYYMMDD. This was used to find the local continuum level via a linear fit. (b) We then performed cosmic ray and telluric line rejections, before (c) estimating the local S/N. (d) We automatically selected a series of flux points contributing to a given absorption line, and then (e) we derived abundances by comparing the observed spectrum with a set of convolved synthetic spectra characterized by different abundances. Subsequently, four different abundance determination methods were used: (1) line-profile fitting; (2) core line-intensity comparison; (3) global goodness-of-fit estimate; and (4) an equivalent-width comparison. Each diagnostic yields validation flags. Based on these flags, a decision tree then rejects or accepts each estimate, keeping the best-fit abundance. We adopted the $\xi^{2}$ diagnostic for the abundance choice because of its robustness. However, we stored the information from the other diagnostics, including the standard deviation between all four methods. 

\begin{figure}
        \centering
        \includegraphics[width=0.5\textwidth]{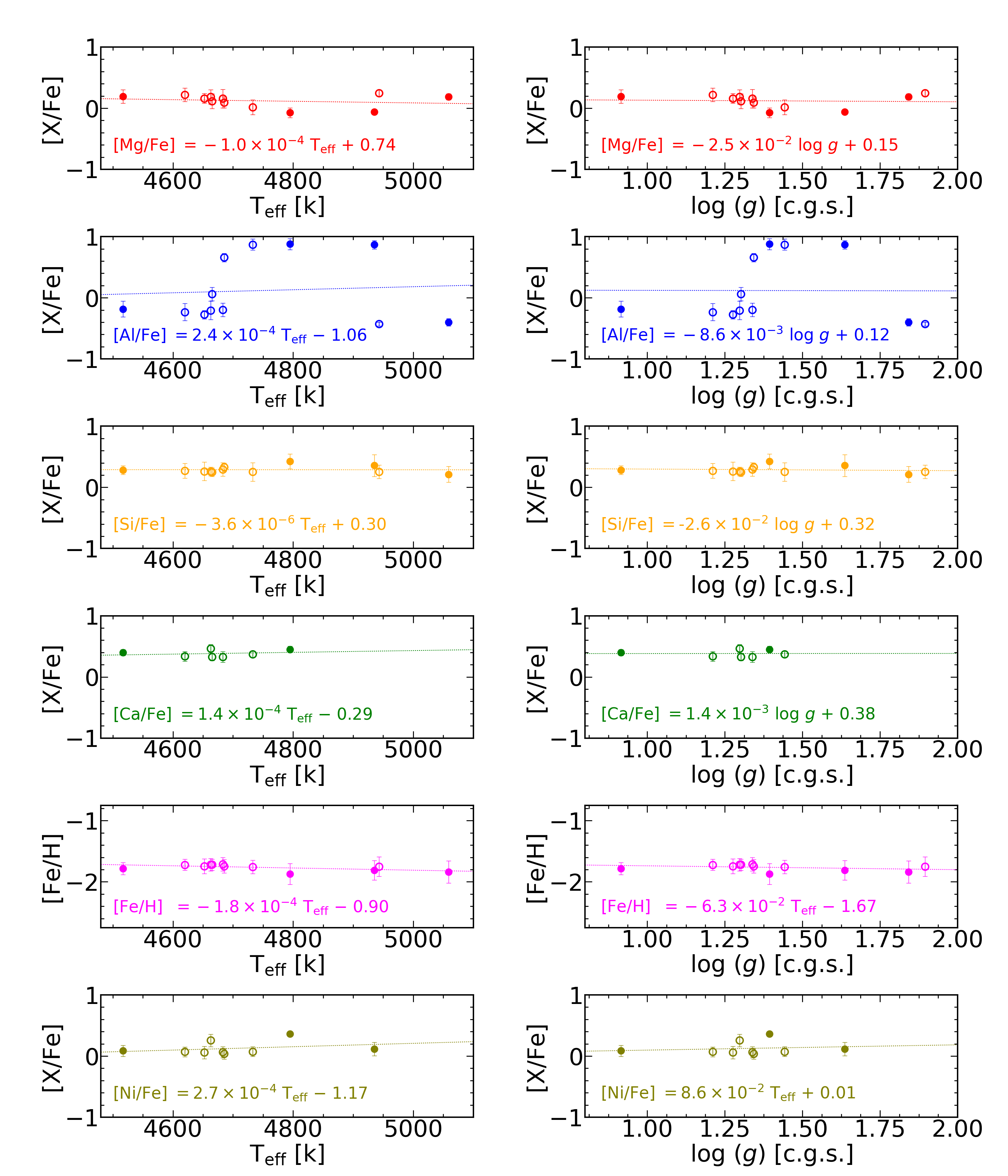}
        \caption{Elemental abundances and uncertainties of $\alpha$ elements (Mg, Si, and Ca), the odd-Z element Al, and Fe-peak elements (Fe and Ni) determined with \texttt{BACCHUS} for stars in NGC~2298, as a function of the atmospheric parameters (T$_{\rm eff}$ and $\log$ \textit{g}). The RGB and AGB stars are denoted by open and filled symbols, respectively. The dashed lines and inset notation represent the linear regression of the data.}
        \label{Figure5}
\end{figure}

\section{Elemental abundances}
\label{Results}

The present study substantially contributes to the chemical characterisation of the metal-poor GC NGC~2298. Our sample of 12 cluster members allows us to investigate any intra-cluster abundance variations. Figure \ref{Figure3} summarizes the resulting elemental abundances for six chemical species, including $\alpha$ elements (Mg, Si, and Ca), the odd-Z element Al, and iron-peak elements (Fe and Ni)  that were investigated from APOGEE-2S spectra for the 12 stars in the innermost region of NGC~2298. Table \ref{table1} contains the abundance values for [Mg/Fe], [Al/Fe], [Si/Fe], [Ca/Fe], [Fe/H], and [Ni/Fe] derived by the \texttt{BACCHUS} code scaled to the Solar reference value from \citet{Asplund2005}. 

We remark that no conclusive answers can be drawn for C, N, and O from $^{12}$C$^{14}$N, $^{16}$OH, and $^{12}$C$^{16}$O, as these lines become too weak to be detected or accurately measured at the metallicity and temperature range of our NGC~2298 stars. For this reason, we choose not to speculate on the interpretation of C, N, and O abundances for the present sample.

Overall, the results for NGC~2298 are in reasonable agreement with other GCs of a similar metallicity for almost all species. However, NGC~2298 exhibits median values for [N/Fe] and [Mg/Fe] which are slightly lower than M~53, M~55, and NGC~6397 \citep{Meszaros2020}, suggesting a likely different chemical-enrichment history and/or distinct birth conditions. 

Figure \ref{Figure5} demonstrates a lack of abundance dependence on the effective temperature or surface gravity, as well as on the evolutionary status (RGB or AGB) for Mg, Al, Si, Ca, Fe, or Ni at the metallicity of NGC~2298.

In the following, we describe the chemical-enrichment level of NGC~2298 for the different chemical species examined in this work. \newline

\subsection{ $\alpha$ elements (Mg, Si, and Ca)}
 
 From inspection of Table \ref{table1}, the median and mean values for the $\alpha$ elements from Mg to Ca in NGC~2298 are overabundant ($\gtrsim +0.12$) compared to the Sun, supporting the idea that the main contributors to the chemistry of this cluster have likely been mostly provided by supernovae (SNe) II events \citep{Tsujimoto2012}, which have been formed before SNe Ia could significantly contribute with iron. This is a feature very common to almost all Galactic GCs with a similar metallicity in the outer halo, and a metallicity below [Fe/H] $=-1.0$.    
 
The NGC~2298 sample has a star-to-star spread in [Si/Fe]$\sim0.15$ dex and [Ca/Fe]$\sim0.13$ dex, which is slightly larger than the  measurement uncertainties for each species, with the exception of [Mg/Fe]. There is a significant [Mg/Fe] spread of $\sim0.3$ dex, exceeding the observational uncertainties, which is anti-correlated with [Al/Fe] (see subsection \ref{aluminum}). However, the [Mg/Fe] abundance ratio by itself exhibits a median value that is systematically offset by roughly 0.2 dex lower than M~53, M~55, and NGC~6397 stars from \citet{Meszaros2020}, as can be appreciated in Figure \ref{Figure3}. There are several reasons for a systematic offset between our [Mg/Fe] determinations and those determined by \citet{Meszaros2020}. For instance, non-local thermodynamic equilibrium (NLTE) and/or 3D effects are currently not modeled when fitting the APOGEE spectra; and/or the systematic differences in the temperature scales.
 
Figure \ref{Figure4} reveals an apparent Mg-Si anti-correlation. The existence of a Mg-Si anti-correlation in NGC~2298 implies the presence of leakage from the MgAl chain into Si production through the $^{26}$Al(p,$\gamma^{27}$Si(e$-$,$\nu$)$^{27}$Al(p,$\gamma$)$^{28}$Si reactions at a high temperature. In the absence of this leakage, we would expect a simple correlation between Mg and Si since they are both $\alpha$ elements \citep[see, e.g.,][]{Yong2005, Carretta2009, Meszaros2020}. We find that some of the stars in NGC~2298 exhibit a higher Si abundance than their Mg-rich counterparts, confirming the likely occurrence of hot proton burning in the early populations of NGC~2298. It is important to note that the evidence for this Mg-Si anti-correlation does not exist for the other GCs at a similar metallicity as that of NGC~2298.

An Al-Si correlation is also observed in NGC~2298, as can be appreciated in Figure \ref{Figure4}. \citet{Yong2005}, \citet{Carretta2009}, \citet{Meszaros2015}, \citet{Masseron2019}, and \citet{Meszaros2020} interpreted the Al-Si correlation as a signature of $^{28}$Si leakage from the Mg-Al chain. Interestingly, most of the Si-enriched stars in NGC~2298 also seem to correspond to the Mg-depleted and Al-enhanced cluster population, which are likely the result of Ne-Na and Mg-Al cycles occurring in the H-burning shell of the first-generation stars whose nucleosynthetic products were later distributed through the cluster. We stress that, regarding any possible analysis bias, we have not been able to find any dependence of the abundances with the evolutionary status.

\subsection{The odd-Z element Al}
\label{aluminum}
 
\begin{figure*}
        \centering
        \includegraphics[width=1.0\textwidth]{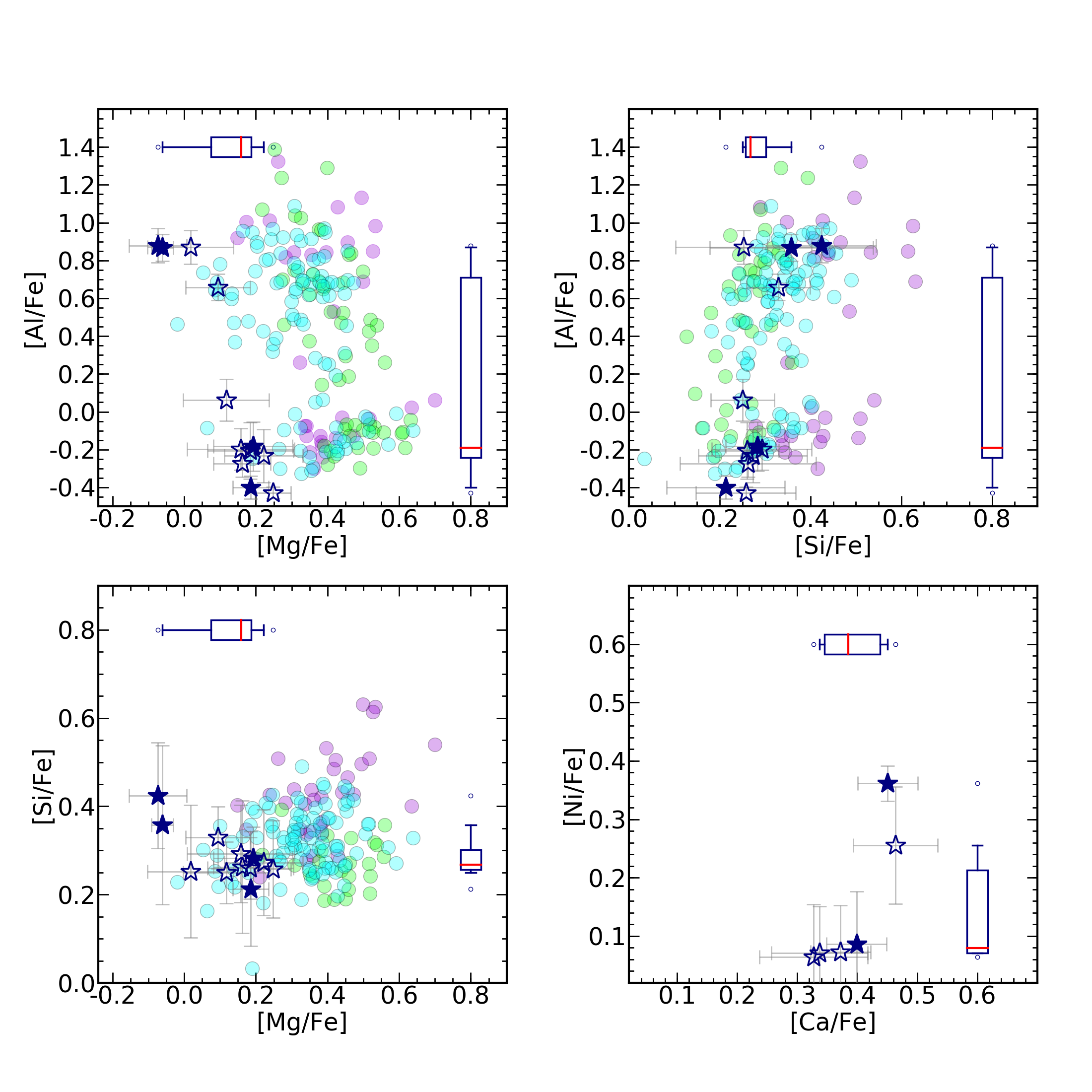}
        \caption{Combined elemental abundances of $\alpha$ elements (Mg, Si, and Ca), the odd-Z element Al, and iron-peak elements (Fe and Ni) compared to cluster stars taken from \citet{Meszaros2020} at a similar metallicity as NGC~2298. Stars in NGC~2298 are denoted by navy blue star symbols, with open and filled symbols indicating RGB and AGB stars, respectively. The inset box plots indicates, with horizontal red lines, the median as well as the 5$^{\rm th}$ and 95$^{\rm th}$ percentile limits of the distribution of each chemical species in NGC~2298. The gray symbols represent the internal uncertainties of each element.}
        \label{Figure4}
 \end{figure*}

Similar to the result reported 30 years ago by \citet{McWilliam1992}, we confirm that this cluster also hosts a clear population of Al-enhanced stars with [Al/Fe] $\gtrsim+0.5$, which is significantly higher than typical Galactic abundances of [Al/Fe]$< +0.5$ (for the typical first-generation stars). Figure \ref{Figure4} also clearly reveals two distinct groups in the Mg-Al plane; one of them lies in the region dominated by the lowest [Al/Fe] abundances with slightly super-Solar [Mg/Fe] abundance ratios rather than sub-Solar ones (often called first-generation stars or unenriched stars), while the second group is governed by higher [Al/Fe] abundances with the lowest [Mg/Fe] abundances, containing a fraction of stars well below [Mg/Fe] $\approx$0 (called second-generation or enriched stars). We conclude that the intra-GC abundance variations reported in Figure \ref{Figure4} are indicative of the presence of the MP phenomenon in NGC~2298, but the origin of the polluting material remains elusive based on these data. 
 
 We also detect a significant star-to-star spread ($\gtrsim1.29$ dex) in [Al/Fe], as can be appreciated in Figure \ref{Figure4}. An anti-correlation between Al and Mg is clearly present in NGC~2298, showing the signs of proton burning of the envelope material by the Mg-Al cycle, which is commonly present in most metal-poor GCs. It is clear that the extended distribution of Al and Mg is much larger than the typical errors of [Al/Fe] and [Mg/Fe]. 
 
 With our cluster sample, we agree with the previous claim by \citet{McWilliam1992} regarding the presence of a strong Al enhancement in NGC~2298.\ However, our large sample confirms, for the first time, the presence of a clear Mg-Al anti-correlation in NGC~2298, whose proton-burning Mg-Al cycle drives the excess of Al through the reduced surface abundances of $^{24}$Mg \citep[see e.g.,][]{Ward1980}. 
 
 We also find that the RGB and AGB stars do not group separately in any of the abundance-abundance planes presented in Figure \ref{Figure4}. Therefore, the observed Mg-Al anti-correlation in NGC~2298 does not depend on the evolutionary status of a star. 

 \subsection{Iron-peak elements (Fe and Ni)}
 
We find a mean metallicity $\langle$[Fe/H]$\rangle =-1.76 \pm 0.01$ with a dispersion of $\sigma_{\rm [Fe/H]} = 0.048\pm0.009$ dex. Reported errors are errors on the mean.  We also find an iron star-to-star spread of 0.14 dex, compatible with the measurement errors, so we have no evidence for an intrinsic Fe abundance spread in NGC~2298. Our measured mean metallicity for this GC exhibits a deviation greater than the uncertainties in comparison to some previous works that employ a variety of methods reporting a wide range ($\sim$0.25 dex) of metallicity from [Fe/H]$ = -1.96$ to $-1.71$, but it is in reasonable agreement with a few of them. For instance, \citet{Frogel1983} transformed optical and NIR colors to Cohen's metallicity scale \citep{Cohen1983}, and they estimated a mean value of [Fe/H]$=-1.76$, which has been listed as [Fe/H]$_{\rm IR}$ in Table 5 of \citet{Zinn1984}; 
\citet{Zinn1984} found a mean metallicity of [Fe/H]$=-1.85$; \citet{Zinn1985} found a [Fe/H]$=-1.81$ on his metallicity scale; \citet{McWilliam1992} found a metallicity of [Fe/H] $= -1.91$; \citet{Geisler1995} estimated a metallicity of [Fe/H] $= -1.82$; \citet{Salaris1996} reported a metallicity of [Fe/H]$=-1.91$; \citet{Carretta1997} reported a metallicity of [Fe/H] $= -1.71$;  while \citet{Kraft2003} listed a range of metallicity [Fe/H] between $-1.93$ to $-1.83$; \citet{Pritzl2005} reported a metallicity of [Fe/H] $= -1.90$;
\citet{Carretta2009a} provided a metallicity of [Fe/H]$-1.96$; \citet{Carrera2013} listed a [Fe/H]$=-1.74$; while more recently \citet{Roediger2014} and \citet{Yong2014} have provided a metallicity estimation of [Fe/H] $= -1.95$ and [Fe/H] $= -1.96$, respectively.

As far as other iron-peak elements are concerned, Ni is slightly overabundant relative to Solar ($+0.12$), with a star-to-star spread of 0.27 dex that is higher than the measurement uncertainties of [Ni/Fe] abundance ratios in NGC~2298, which strikingly appear to be weakly correlated with [Ca/Fe], as can be appreciated in Figure \ref{Figure4}. However, the observed star-to-star spread in the [Ca/Fe] abundance ratios is not statistically significant according to our error analysis, as has been observed in other GCs \citep{Carretta2021}. Our sample is tightly concentrated around the mean [Ca/Fe] of NGC~2298. Therefore, with the determinations of [Ni/Fe] and [Ca/Fe] for a limited sample of six cluster stars, we cannot draw firm conclusions about the apparent correlation between these two species.  This possibility will have to be considered using larger future samples.

\section{Concluding remarks}
\label{conclusions}

We have performed a high-resolution spectral analysis for 12 stars in the old GC NGC~2298. This cluster is located in a region of low interstellar reddening, and it has a halo orbit that crosses the Bulge, with an apocentric distance of $\sim$16.44 kpc \citep{Massari2019}. We found a mean and median metallicity of [Fe/H] $=-1.76$ and $-1.75$, respectively, with a star-to-star spread of 0.14 dex that is compatible with the measurement errors, so we find no evidence for an intrinsic Fe abundance spread in NGC~2298. Our reported metallicity is $\sim0.2$ dex more metal-rich than previously thought, but it is in reasonable agreement with estimations provided by \citet{Kraft2003} and \citet{Carrera2013}. 
        
We confirm the existence of an Al-enriched population in NGC~2298, as was claimed three decades ago. We provide, for the first time, evidence for the standard anti-correlation between Mg and Al in our data. This result indicates the prevalence of the MPs phenomenon at the low metallicity of NGC~2298. It is also important to note that RGB and AGB stars present in NGC~2298 do not appear to follow different paths or group separately in any of the abundance planes examined in this work; therefore, we conclude that the observed Mg-Al anti-correlation does not depend on the evolutionary status of a star in NGC~2298. We also detect an apparent Mg-Si anti-correlation and an Al-Si correlation, which are likely signatures of $^{28}$Si leakage from the Mg-Al chain, which is a feature common to other GCs \citep[][]{Masseron2019, Meszaros2020} at a similar metallicity as NGC~2298.  

\begin{acknowledgements}  
We thank the useful discussions and comments from Baitian Tang, Christian Nitschelm, and an anonymous referee to improve the manuscript.\\ 
J.G.F-T gratefully acknowledges the grant support provided by Proyecto Fondecyt Iniciaci\'on No. 11220340, from ANID Concurso de Fomento a la Vinculaci\'on Internacional para Instituciones de Investigaci\'on Regionales (Modalidad corta duraci\'on) Proyecto No. FOVI210020, and also from the grant support from the Joint Committee ESO-Government of Chile 2021 (ORP 023/2021).\\
S.V. gratefully acknowledges the support provided by Fondecyt regular n. 1220264, and by the ANID BASAL projects ACE210002 and FB210003.\\
T.C.B. acknowledges partial support for this work from grant PHY 14-30152: Physics Frontier Center / JINA Center for the Evolution of the Elements (JINA-CEE), awarded by the US National Science Foundation.\\
B.B. acknowledges grants from FAPESP, CNPq and CAPES - Financial code 001.\\ 
D.M. gratefully acknowledges support by the ANID BASAL projects ACE210002 and FB210003, and Fondecyt Project No. 1220724.\\
E.R.G acknowledges support from ANID PhD scholarship No. 21210330.\\
D.G. gratefully acknowledges support from the ANID BASAL project ACE210002.\\
D.G. also acknowledges financial support from the Direcci\'on de Investigaci\'on y Desarrollo de la Universidad de La Serena through the Programa de Incentivo a la Investigaci\'on de Acad\'emicos (PIA-DIDULS).\\

Funding for the Sloan Digital Sky Survey IV has been provided by the Alfred P. Sloan Foundation, the U.S. Department of Energy Office of Science, and the Participating Institutions. SDSS- IV acknowledges support and resources from the Center for High-Performance Computing at the University of Utah. The SDSS web site is www.sdss.org. SDSS-IV is managed by the Astrophysical Research Consortium for the Participating Institutions of the SDSS Collaboration including the Brazilian Participation Group, the Carnegie Institution for Science, Carnegie Mellon University, the Chilean Participation Group, the French Participation Group, Harvard-Smithsonian Center for Astrophysics, Instituto de Astrof\`{i}sica de Canarias, The Johns Hopkins University, Kavli Institute for the Physics and Mathematics of the Universe (IPMU) / University of Tokyo, Lawrence Berkeley National Laboratory, Leibniz Institut f\"{u}r Astrophysik Potsdam (AIP), Max-Planck-Institut f\"{u}r Astronomie (MPIA Heidelberg), Max-Planck-Institut f\"{u}r Astrophysik (MPA Garching), Max-Planck-Institut f\"{u}r Extraterrestrische Physik (MPE), National Astronomical Observatory of China, New Mexico State University, New York University, University of Notre Dame, Observat\'{o}rio Nacional / MCTI, The Ohio State University, Pennsylvania State University, Shanghai Astronomical Observatory, United Kingdom Participation Group, Universidad Nacional Aut\'{o}noma de M\'{e}xico, University of Arizona, University of Colorado Boulder, University of Oxford, University of Portsmouth, University of Utah, University of Virginia, University of Washington, University of Wisconsin, Vanderbilt University, and Yale University.\\

This work has made use of data from the European Space Agency (ESA) mission \textit{Gaia} (\url{http://www.cosmos.esa.int/gaia}), processed by the \textit{Gaia} Data Processing and Analysis Consortium (DPAC, \url{http://www.cosmos.esa.int/web/gaia/dpac/consortium}). Funding for the DPAC has been provided by national institutions, in particular the institutions participating in the \textit{Gaia} Multilateral Agreement.\\
\end{acknowledgements}

%\bibliographystyle{aa}
%\bibliography{references}

\begin{appendix}
        
\section{Basic parameters of NGC~2298 stars}

The basic parameters for the 12 NGC~2298 members examined in this work are listed in Table \ref{table1}.

\begin{sidewaystable*}
        %\begin{table*}
        \begin{small}
                \begin{center}
                        \setlength{\tabcolsep}{1.7mm}  
                        \caption{\textit{Gaia} EDR3 photometric and astrometric, APOGEE-2S radial velocity, photometric atmospheric (T$_{\rm eff}$, $\log$ (\textit{g}) and $\xi_{t}$) properties, and \texttt{BACCHUS} elemental abundances for 12 stars in the innermost regions of NGC~2298. The cluster median, mean, and standard deviation from \texttt{BACCHUS} and \texttt{ASPCAP}, as well as the \texttt{BACCHUS} spread are listed in the last rows.}
                        \begin{tabular}{|lccccccccccccccccc|}
                                \hline
                                APOGEE\_ID            &  Class & G     &  B$_{\rm p}$     &  R$_{\rm p}$    &  $\mu_{\alpha} \cos{}(\delta) \pm \Delta$          &     $\mu_{\delta} \pm \Delta$  &   RV     & S/N  & T$_{\rm eff}$ & $\log$ (\textit{g}) &  $\xi_{t}$       &   [Mg/Fe]     &   [Al/Fe]     &  [Si/Fe]    &  [Ca/Fe]   & [Fe/H]      &  [Ni/Fe]        \\  
                                &     & &       &   & mas yr$^{-1}$          &    mas yr$^{-1}$    &    km s$^{-1}$    &   pixel$^{-1}$   & K & [cgs] &  km s$^{-1}$   &         &      &    &     &      &        \\  
                                \hline
                                \hline
                                2M06485459$-$3559151 & RGB & 13.89 &  14.61  &  13.06 &  3.18$\pm$0.01 &   $-$2.21$\pm$0.01 & 147.35 &  156  & 4685 & 1.34 &  1.95  &    $+$0.09   &     $+$0.65 &  $+$0.32  &  ...  & $-$1.74 &  $+$0.03 \\
                                &  &  &  &  &  &  &  &  &  &  &  &  (0.09) &  (0.07) &  (0.07) & ...  & (0.11) &  (0.09) \\
                                2M06485872$-$3600006 & RGB & 13.75 &  14.43  &  12.92 &  3.28$\pm$0.01 &   $-$2.18$\pm$0.01 & 147.61 &  174  & 4652 & 1.27 &  1.98  &    $+$0.16   &  $-$0.27 &  $+$0.26  &  ...  & $-$1.74 &  $+$0.05 \\
                                &  &  &  &  &  &  &  &  &  &  &  &  (0.08) &  (0.07) &  (0.15) & ... & (0.12) &  (0.10) \\
                                2M06485886$-$3558262 & RGB & 13.87 &  14.57  &  13.06 &  3.35$\pm$0.01 &   $-$2.14$\pm$0.01 & 146.71 &  166  & 4683 & 1.33 &  1.96  &    $+$0.15   &  $-$0.19 &  $+$0.29  &  $+$0.32 & $-$1.71 &  $+$0.06 \\     
                                &  &  &  &  &  &  &  &  &  &  &  &  (0.15) &  (0.11) &  (0.11) & (0.09) & (0.11) &  (0.09) \\
                                2M06485916$-$3600503 & RGB & 13.80 &  14.50  &  12.97 &  3.33$\pm$0.01 &   $-$2.25$\pm$0.01 & 142.10 &  169  & 4663 & 1.29 &  1.97  &    $+$0.18   &  $-$0.20 &  $+$0.25  &  $+$0.46 & $-$1.72 &  $+$0.25 \\   
                                &  &  &  &  &  &  &  &  &  &  &  &  (0.12) &  (0.15) &  (0.07) & (0.07) & (0.10) &  (0.10) \\
                                2M06490301$-$3559100 & RGB & 13.80 &  14.50  &  12.98 &  3.34$\pm$0.01 &   $-$2.13$\pm$0.01 & 145.75 &  157  & 4665 & 1.30 &  1.97  &    $+$0.11   &     $+$0.06 &  $+$0.24  &  $+$0.32 & $-$1.71 &  ...  \\         
                                &  &  &  &  &  &  &  &  &  &  &  &  (0.12) &  (0.11) &  (0.07) & (0.06) & (0.09) &  ... \\
                                2M06490382$-$3600299 & RGB &  14.97 &  15.56  &  14.20 &  3.40$\pm$0.02 &   $-$2.23$\pm$0.02 & 152.96 &  109  & 4943 & 1.89 &  1.75  &    $+$0.24   &  $-$0.42 &  $+$0.25  &  ...  & $-$1.75 &  ...  \\            
                                &  &  &  &  &  &  &  &  &  &  &  &  (0.05) &  (0.03) &  (0.11) & ...  & (0.16) &  ...  \\
                                2M06490734$-$3558013 & RGB & 13.63 &  14.35  &  12.81 &  3.30$\pm$0.01 &   $-$2.13$\pm$0.01 & 144.48 &  190  & 4620 & 1.21 &  2.00  &    $+$0.22   &  $-$0.23 &  $+$0.27  &  $+$0.33 & $-$1.72 &  $+$0.07 \\   
                                &  &  &  &  &  &  &  &  &  &  &  &  (0.11) &  (0.14) &  (0.12) & (0.08) & (0.09) &  (0.08) \\   
                                2M06491677$-$3557505 & RGB & 14.07 &  14.72  &  13.28 &  3.31$\pm$0.01 &   $-$2.20$\pm$0.01 & 148.22 &  139  & 4733 & 1.44 &  1.92 &    $+$0.01   &    $+$0.87 &  $+$0.25  &  $+$0.37 & $-$1.75 &  $+$0.07 \\  
                                &  &  &  &  &  &  &  &  &  &  &  &  (0.12) &  (0.09) &  (0.15) & (0.05) & (0.11) &  (0.08) \\        
                                2M06485140$-$3600380 & AGB & 13.93 &  14.55  &  13.17 &  3.27$\pm$0.01 &   $-$2.06$\pm$0.01 & 147.95 &  138  & 4795 & 1.39 &  2.03   & $-$0.07   &  $+$0.87 &  $+$0.42  &  $+$0.45 & $-$1.87 &  $+$0.36 \\       
                                &  &  &  &  &  &  &  &  &  &  &  &  (0.08) &  (0.09) &  (0.12) & (0.05) & (0.17) &  (0.03) \\
                                2M06485619$-$3602261 & AGB & 14.78 &  15.32  &  14.06 &  3.37$\pm$0.01 &   $-$2.19$\pm$0.01 & 150.23 &  112  & 5059 & 1.84 &  1.90  &    $+$0.18   &  $-$0.40 &  $+$0.21  &  ...  & $-$1.83 &  ...  \\       
                                &  &  &  &  &  &  &  &  &  &  &  &  (0.05) &  (0.06) &  (0.13) & ...  & (0.18) &  ... \\
                                2M06485888$-$3601165 & AGB & 13.09 &  13.85  &  12.22 &  3.38$\pm$0.01 &   $-$2.18$\pm$0.01 & 144.04 &  254  & 4517 & 0.91 &  2.17  &    $+$0.19   &  $-$0.18 &  $+$0.28  &  $+$0.39 & $-$1.78 &  $+$0.08 \\   
                                &  &  &  &  &  &  &  &  &  &  &  &  (0.11) &  (0.13) &  (0.07) & (0.05) & (0.10) &  (0.09) \\
                                2M06485948$-$3559426 & AGB & 14.40 &  14.93  &  13.58 &  3.43$\pm$0.01 &   $-$2.24$\pm$0.01 & 141.92 &  116  & 4935 & 1.63 &  1.95 & $-$0.06   &  $+$0.86 &  $+$0.35  &  ...  & $-$1.80 &  $+$0.11 \\
                                &  &  &  &  &  &  &  &  &  &  &  &  (0.03) &  (0.07) &  (0.18) & ...  & (0.16) &  (0.11) \\
                                \hline
                                \hline
                                \texttt{BACCHUS} &      &       &       &       &       &       &       &       &       &       &       &  &  &  & &  &  \\               
                                \hline                          
                                Cluster median       & ... & ...  & ...  & ... &...  & ... & ... & ... &...  & ... & ... & $+0.16$ & $-0.19$ & $+0.27$ & $+0.37$& $-1.75$ & $+0.07$ \\                 
                                Cluster mean           & ... & ...  & ...  & ... &...  & ... & ... & ... &...  & ... & ... &  $+0.12$ & $+0.12$ & $+0.29$ & $+0.38$ & $-1.76$ & $+0.12$\\                 
                                Std                               & ... & ...  & ...  & ... &...  & ... & ... & ... &...  & ... & ... & $0.10$ & $0.51$ & $0.05$ & $0.05$& $0.05$ & $0.10$ \\                   
                                Spread$\dagger{}$  & ... & ...  & ...  & ... &...  & ... & ... & ... &...  & ... & ...  & $0.30$ & $1.29$ & $0.15$ & $0.13$& $0.14$ & $0.27$ \\                    
                                \hline
                        \end{tabular}  \label{table1}
                \end{center}
                \raggedright{$\dagger{}$Spread is defined as the $95^{\rm th}$ percentile - $5^{\rm th}$percentile. The listed uncertainty inside parentheses for each chemical species is as follows: $\sigma_{total}  = \sqrt{\sigma^2_{T_{\rm eff}}    + \sigma^2_{{\rm \log} \textit{g}} + \sigma^2_{\xi_t}  + \sigma^2_{mean}  }$, where $\sigma^2_{mean} $ was calculated using the standard deviation from the different abundances of the different lines for each element, while  $\sigma^2_{\rm T_{eff}} $ ,  $\sigma^2_{{\rm \log} \textit{g}}$, and $\sigma^2_{\xi_t} $ were derived for each chemical species while varying T$_{\rm eff}$ by $\pm$100 K, $\log$ \textit{g} by $\pm0.3$ dex, and $\xi_{t}$ by $\pm0.05$ km s$^{-1}$. }       
        \end{small}
        %\end{table*}   
\end{sidewaystable*}    
                
\end{appendix} 

\end{document}